\documentclass[aps,showkeys,twocolumn,pra]{revtex4}
\usepackage{amsmath,amsthm,amsfonts,amssymb,amsbsy,mathrsfs}
\usepackage{units,times,float,graphicx,bm,comment}
\begin{document}
\title{Quantum acceleration by an ancillary system in non-Markovian environments}
\author{Jinna Fan$^1$}
\author{Shaoxiong Wu$^1$\footnote{sxwu@nuc.edu.cn}}
\author{Chang-shui Yu$^2$\footnote{ycs@dlut.edu.cn}}
\affiliation{$^1$ School of Science, North University of China, Taiyuan 030051, China\\
$^2$ School of Physics, Dalian University of Technology, Dalian 116024, China}
\date{\today }

\begin{abstract}
We study the effect of an ancillary system on the quantum speed limit time in different non-Markovian environments. Through employing an ancillary system coupled with the quantum system of interest via hopping interaction and investigating the cases that both the quantum system and ancillary system interact with their independent/common environment, and the case that only the system of interest interacts with the environment, we find that the quantum speed limit time will become shorter with enhancing the interaction between the system and environment and show periodic oscillation phenomena along with the hopping interaction between the quantum system and ancillary system increasing. The results indicate that the hopping interaction with the ancillary system and the structure of environment determine the degree of which the evolution of the quantum system can be accelerated.
\end{abstract}
\keywords{Quantum speed limit, Open quantum systems, Hopping interaction, Ancillary system}
\maketitle

\section{Introduction}

The Heisenberg uncertainty relationship is one of the features that distinguish the quantum mechanics from the classical physics. For the closed quantum systems, the dynamics of the pure state $|\psi \rangle $ is governed by the Schr\"{o}dinger equation. What is the shortest time for the pure state evolving to its orthogonal state? In order to answer this question, Mandelstam and Tamm proposed the notion of (MT) quantum speed limit time (QSLT) in 1945 \cite{Mandelstam45}, which is related to the variance of the energy $\frac{\pi \hbar }{2\Delta E}$ and can be considered as the extension of the Heisenberg time-energy relation. According to the transition probability amplitude between two orthogonal quantum pure states $\langle \psi _{0}|\psi _{\tau }\rangle $, Margolus and Levitin \cite{Margolus98} presented another form (ML) of quantum speed limit time in connection with the mean value of energy $\frac{\pi\hbar }{2(\langle E\rangle-E_0)}$. Combined the MT with ML bounds  of QSLT, the unified bound of quantum speed limit time can be given as $\tau _{\text{qsl}}=\max \{\frac{\pi \hbar }{2\Delta E},\frac{\pi \hbar }{2(\langle E\rangle-E_0)}\}$. The quantum speed limit had been investigated in different ways for closed quantum systems \cite{Levitin09,Fleming73,Bhattacharyya83,Anandan90,Pati91,Vaidman92,Brody03,Jones10,Campaioli18}, it can deepen the understanding of the uncertainty relationship, the evolution of quantum state, the quantum control feedback \cite{Caneva09}, and even the information of black hole \cite{Bekenstein81}.

With the numerous research on the open quantum systems \cite{Breuer16}, the quantum speed limit had been extended to the non-Markovian open quantum systems in recent years. In Ref. \cite{Taddei13}, the authors derived an attainable lower bound for time estimation by connecting with the quantum Fisher information, and it can be considered as a MT bound. In Ref. \cite{Campo13}, the authors derived a time-energy uncertainty relation for the open quantum system under completely positive and trace preserving evolution, whose relation is analogous to MT bound. In Ref. \cite{Deffner13}, utilizing the von Neumann trace inequality and Cauchy-Schwarz inequality for operators, the authors obtained a unified form of quantum speed limit time for the open quantum systems, and discussed the effects of Markovianity. Since then, the quantum speed limit of open quantum systems had attracted widespread attentions \cite{Frey16,Deffner17jpa}, such as the property of the quantum speed limit in non-Markovian environments \cite{Xu14,Zhang14,Liu16,Wu15,Wu18,Song16,Zhang15,Sun15,Ekt17,Wu20sr,Awasthi20,Jing16}, the connection between the quantum speed limit and the skew information \cite{Pires16,Mar16} or entropy \cite{Deffner19,Campaioli20}, the application in shortcuts to adiabaticity \cite{Campbell17,Xu18,Demirplak08,Takahashi13,Campo14,Funo17}, the distinct bound with gauge invariant distance \cite{Sun19}, the relation with the condensed matter physics \cite{Fogarty20,Puebla20}, and many other aspects \cite{GP19,Nicholson20}. Recently, the speed limit in the phase space have also been reported \cite{Shanahan18,Okuyama18,Wu20cpb,Hu20,Shiraishi18,Deffner17njp}. Most works are mainly focused on the effects of environments on the QSLT or the fantastic properties of the QSLT by employing different ``distance" measurements. However, the effects of ancillary system on the quantum speed limit time is still an open question.

In this paper, we will consider a two-level quantum system (``system", for short) coupled with an ancillary system (``ancilla", for short) through the hopping interaction $\hbar J(\sigma _{a}^{+}\sigma_{b}^{-}+\sigma _{a}^{-}\sigma _{b}^{+})$, and investigate the evolution of the system in different environments.  We mainly study the follow three cases: the first case is that the system and the ancilla interact with independent environments, respectively; the second case is that both the system and the ancilla interact with common environment; and the third case is that only the system of interest interacts with environment. Without loss of generality, we assume that the system and the ancilla are the same, and the structure of environment is assumed as the Lorentzian form. We find that the quantum speed limit time can be decreased with the increasing of the coupling strength $\gamma _{0}$ between the system-ancilla and environment. Meanwhile, with the enhancement of hopping interaction $J$ between the system and the ancilla, the quantum speed limit time will become shorter, and show periodic oscillation phenomena. The hopping interaction between the system and the ancilla and the structure of environment determine the degree that the evolution of quantum state can be accelerated.

This paper is organized as follows. In Sec. II, an introduction about the quantum speed limit time of open quantum systems is given. In Sec. III, we consider the QSLT for the coupled system-ancilla interacting with independent environments. In Sec. IV, we consider the QSLT for the coupled system-ancilla interacting with common environment. In Sec. V, the QSLT is studied when only the system of interest interacts with environment. At the end, the conclusion and discussion are given.

\section{The quantum speed limit of open quantum systems}

We will start with a brief introduction about the quantum speed limit time of open quantum systems following the Ref. \cite{Deffner13}. The fidelity between the initial state $\rho_0$ and the final state $\rho_{\tau}$ is $F(\rho _{0},\rho _{\tau})=(\text{tr}[\sqrt{\sqrt{\rho _{0}}\rho _{\tau }\sqrt{\rho _{0}}}])^{2}$, and the geometric distance measure between quantum states is chosen as Bures angle $\mathcal{L}(\rho _{0},\rho _{\tau })=\arccos [\sqrt{F(\rho _{0},\rho_{\tau })}]$. When the initial state is a pure state $|\psi _{0}\rangle$, the Bures angle can be simplified as
\begin{equation}
\mathcal{L}(\rho _{0},\rho _{\tau })=\arccos [\sqrt{\langle \psi _{0}|\rho_{\tau }|\psi _{0}\rangle }].
\end{equation}

Utilizing the von Neumann trace inequality for operators, one can obtain the Margolus-Levitin bound of quantum speed limit time of open quantum systems
\begin{equation}
\tau \geq \max \left\{ \frac{1}{\Lambda _{\tau }^{\text{op}}},\frac{1}{\Lambda _{\tau }^{\text{tr}}}\right\} \sin ^{2}[\mathcal{L}(\rho ,\rho_{\tau })],
\end{equation}%
where $\Lambda _{\tau }^{\text{op,tr}}=(1/\tau )\int_{0}^{\tau }dt\Vert L_{t}(\rho _{t})\Vert _{\text{op,tr}}$ denotes the quantum evolution rate of the open quantum systems, $L_{t}(\rho _{t})$ is the time-dependent nonunitary dynamical operator, and $\Vert \cdot \Vert _{\text{op},\text{tr}}$ means the operator norm and trace norm for the matrix, respectively. Using the Cauchy-Schwarz inequality for operators, one can obtain the Mandelstam-Tamm bound of quantum speed limit time of open quantum systems
\begin{equation}
\tau \geq \frac{1}{\Lambda _{\tau }^{\text{hs}}}\sin ^{2}[\mathcal{L}(\rho,\rho _{\tau })],
\end{equation}%
where $\Lambda _{\tau }^{\text{hs}}=(1/\tau )\int_{0}^{\tau} dt\Vert L_{t}(\rho_{t})\Vert _{\text{hs}}$, and $\Vert \cdot \Vert _{\text{hs}}$ means the Hilbert-Schmidt norm for matrix. Combining the MT and ML bounds, one can obtain a unified form of quantum speed limit time as following
\begin{equation}
\tau _{\text{qsl}}=\max \left\{ \frac{1}{\Lambda _{\tau }^{\text{op}}},\frac{1}{\Lambda _{\tau }^{\text{tr}}},\frac{1}{\Lambda _{\tau }^{\text{hs}}}\right\} \sin ^{2}[\mathcal{L}(\rho _{0},\rho _{\tau })].\label{eq:inqslty}
\end{equation}

According to the matrix theory \cite{Horn85}, the norms of matrix $A$ satisfy the inequality $\Vert A\Vert _{\text{op}}\leq \Vert A\Vert _{\text{hs}}\leq \Vert A\Vert _{\text{tr}}$. It is easy to prove that the ML bound based on the operator norm provides the sharpest bound of quantum speed limit time of open quantum systems. In the following,  we will consider the models that coupled two-level system-ancilla interact with three different non-Markovian environments and investigate the effects of auxiliary system on the quantum speed limit time for the system of interest based on formula (\ref{eq:inqslty}).

\section{The coupled two-level system-ancilla interacting with independent environments}

We will first consider the case that the coupled two-level system-ancilla interact with independent environments. The first system is of interest to investigated (abbreviated by subscript $a$, and short for ``system"), and the second system is the ancillary system (abbreviated by subscript $b$, and short for ``ancilla"). Here, we suppose that both the system and the ancilla are the same, the hopping interaction strength between the system and ancilla is $J$, and the interaction between the system-ancilla and environment is expressed as $g_{k}$. The whole Hamiltonian of the system-ancilla and environments is
\begin{equation}
H_{\text{id}}=H_{a}+H_{b}+H_{ab},
\end{equation}%
where $H_{a} =\hbar \omega _{0}\sigma _{a}^{+}\sigma _{a}^{-}+\sum_{k}\hbar \omega_{k}a_{k}^{\dag }a_{k}+\sum_{k}\hbar (g_{k}^{\ast }\sigma _{a}^{+}a_{k}+\text{h.c.})$ are the Hamiltonian of the quantum system interacting with the independent environment, $H_{b} =\hbar \omega _{0}\sigma _{b}^{+}\sigma _{b}^{-}+\sum_{k}\hbar \omega_{k}b_{k}^{\dag }b_{k}+\sum_{k}\hbar (g_{k}^{\ast }\sigma _{b}^{+}b_{k}+\text{h.c})$ are the Hamiltonian of the ancillary system interacting with the independent environment, and $H_{ab} =\hbar J(\sigma _{a}^{+}\sigma _{b}^{-}+\sigma _{a}^{-}\sigma_{b}^{+})$ is the interaction Hamiltonian between the quantum system and the ancillary system.

The initial state is assumed as follows: the system is in the excited state $|e\rangle $, the ancilla is in the ground state $|g\rangle $, and the independent environments are both in the vacuum states $|00\rangle $. The evolved state in time $t$ is
\begin{align}
|\psi _{\text{id}}(t)\rangle =& A_{\text{id}}(t)|eg\rangle |00\rangle+\sum_{k}C_{\text{id}}^{k}(t)|gg\rangle |1_{k}0\rangle   \notag \\
& +B_{\text{id}}(t)|ge\rangle |00\rangle +\sum_{k}D_{\text{id}}^{k}(t)|gg\rangle |01_{k}\rangle .  \label{eq:2qslpsit}
\end{align}
In the quantum dynamical process, we only consider the evolution of the system (i.e., the first system). The reduced density matrix of the system is
\begin{equation}
\rho _{a}(t)=\left(\begin{array}{cc}
|A_{\text{id}}(t)|^{2} & 0 \\
0 & 1-|A_{\text{id}}(t)|^{2}
\end{array}\right) .\label{eq:2rho}
\end{equation}

According to the non-unitary dynamics of the system $L_{t}(\rho_{a}(t))=\dot{\rho}_{a}(t)$ and the formula (\ref{eq:inqslty}), the tightest bound of quantum speed limit time is the ML bound based on the operator norm
\begin{align}
\tau _{\text{id}}^{\text{qsl}}& =\frac{\sin ^{2}[\mathcal{L}(\rho _{0},\rho_{\tau })]}{\Lambda _{\tau }^{\text{op}}}  \notag \\
& =\frac{1-|A_{\text{id}}(\tau )|^{2}}{\frac{1}{\tau }\int_{0}^{\tau }2|\text{Re}[\dot{A}_{\text{id}}(t)A_{\text{id}}^{\ast }(t)]|dt},
\label{eq:2qsl}
\end{align}%
where $A_{\text{id}}^{\ast }(t)$ is the complex conjugate of coefficient $A_{\text{id}}(t)$ for $|\psi _{\text{id}}(t)\rangle $ in Eq. (\ref{eq:2qslpsit}), and $\dot{A}_{\text{id}}(t)$ is the derivative of $A_{\text{id}}(t)$. The structure of environment interacting with system-ancilla is assumed as the Lorentzian form
\begin{equation*}
I(\omega )=\frac{1}{2\pi }\frac{\gamma _{0}\lambda ^{2}}{(\omega -\omega _{0})^{2}+\lambda^{2}},
\end{equation*}%
where $\omega _{0}$ is the transition frequency of the system, the spectral width parameter $\lambda =1/\tau _{E}$ is related with the environment correlation time, and the coupling strength parameter $\gamma _{0}=1/\tau _{S}$ is connected to the time scales of system change.

Substituting the quantum state $|\psi _{\text{id}}(t)\rangle $ into the Schr\"{o}dinger equation $i\hbar \partial _{t}|\dot{\psi}_{\text{id}}(t)\rangle =H|\psi _{\text{id}}(t)\rangle $, one can obtain the differential equations for the coefficients $A_{\text{id}}(t)$, $B_{\text{id}}(t)$, $C_{\text{id}}^{k}(t)$ and $D_{\text{id}}^{k}(t)$ \cite{Wu18aop}:
\begin{subequations}
\begin{align}
i\dot{A}_{\text{id}}(t)& =\omega _{0}A_{\text{id}}(t)+JB_{\text{id}}(t)+\sum_{k}g_{k}C_{\text{id}}^{k}(t), \\
i\dot{B}_{\text{id}}(t)& =\omega _{0}B_{\text{id}}(t)+JA_{\text{id}}(t)+\sum_{k}g_{k}D_{\text{id}}^{k}(t), \\
i\dot{C}_{\text{id}}^{k}(t)& =\omega _{k}C_{\text{id}}^{k}(t)+g_{k}^* A_{\text{id}}(t), \\
i\dot{D}_{\text{id}}^{k}(t)& =\omega _{k}D_{\text{id}}^{k}(t)+g_{k}^* B_{\text{id}}(t).
\end{align}
\end{subequations}
Using the Laplace transformation and its inverse, the coefficients $A_{\text{id}}(t)$ in the quantum speed limit time (\ref{eq:2qsl}) can be solved analytically as
\begin{equation}
A_{\text{id}}(t)=a_{\text{id}}(t)e^{-i\omega _{0}t},
\end{equation}
where $a_{\text{id}}(t)=\frac{h_{\text{id}}}{2}e^{-\frac{1}{2}(\lambda+iJ)t}+\frac{h_{\text{id}}^{\ast }}{2}e^{-\frac{1}{2}(\lambda -iJ)t}$ with the coefficients $h_{\text{id}}=\cosh [\frac{1}{2}{d_{\text{id}}t}]+\frac{\lambda -iJ}{d_{\text{id}}}\sinh [\frac{1}{2}{d_{\text{id}}t}]$ and $d_{\text{id}}=\sqrt{-J^{2}-2iJ\lambda +\lambda (\lambda-2\gamma _{0} )}$.

\begin{figure}[t]
\centering
\includegraphics[width=0.9\columnwidth]{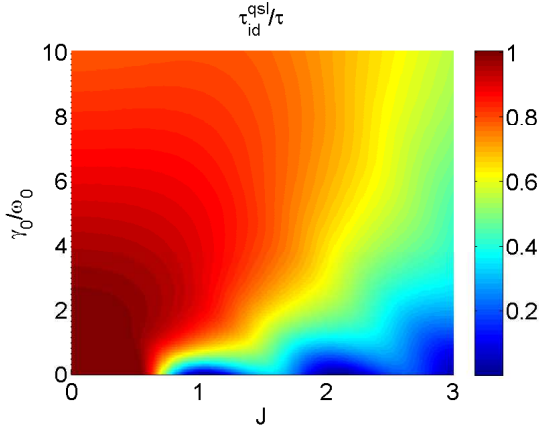}
\caption{The ratio $\tau_{\text{id}}^{\text{qsl}}/\tau$ as the function of system-ancilla hopping interaction parameter $J$ and environment coupling parameter $\gamma_0$. The coupled system-ancilla interact with independent environment, the spectral width parameter is chosen as $\lambda=2$ (in units of $\omega_0$) and the actual driving time is $\tau=3$.}\label{fig1}
\end{figure}

For the ML bound quantum speed limit time $\tau _{\text{id}}^{\text{qsl}}$ in Eq. (\ref{eq:2qsl}), one can find that it not only relates to the coupling strength $\gamma _{0}$ between the system-ancilla and environment, but also is determined by the hopping interaction $J$ between the system and the ancilla. In Fig. \ref{fig1}, we plot the ratio between the quantum speed limit time and the actual driving time $\tau _{\text{id}}^{\text{qsl}}/\tau $ as the function of the environment coupling strength parameter $\gamma _{0}$ and the hopping interaction parameter $J$. The spectral width parameter is chosen as $\lambda =2$ (in units of $\omega _{0}$), and the actual driven time is $\tau =3$.

With the increasing of coupling strength $\gamma _{0}$, the quantum speed limit time shows a decreasing trend, while it demonstrates the phenomenon of periodic oscillation along with the hopping interaction $J$ increasing. Through the hopping interaction $\hbar J(\sigma _{a}^{+}\sigma_{b}^{-}+\sigma _{a}^{-}\sigma _{b}^{+})$, the excited state of the system can be decayed to the ground state and the ancilla can be stimulated to the excited state from the ground state, while the inverse process can also be occurred with probability. It is easy to find that the periodic effect of $J$ on the population of excited state, i.e., $|A_{\text{id}}(t)|^2$, is similar as shown in Fig. \ref{fig1}. Because the non-Markovian interaction between the system-ancilla and environment, there is information flow from the environment back into the system. The comprehensive effects of environment coupling strength and the hopping interaction between the system and the ancilla governs the degree of which the evolution of quantum state can be accelerated. This is a novel phenomenon different from the result of the damped Jaynes-Cummings model \cite{Deffner13}. If the hopping interaction between the system and the ancilla is switched off, i.e., $J=0$, the dynamics of the system and the ancilla are independently and the quantum speed limit time (\ref{eq:2qsl}) will reduce to the result in the damped Jaynes-Cummings model.

\section{The coupled two-level system-ancilla interacting with common environment}

In this section, we will consider the coupled two-level system-ancilla interact with common environment, what different properties of the quantum speed limit exhibit? For this moel, the Hamiltonian is
\begin{equation}
H_{\text{c}}=H_{0}+H_{\text{I}}+H_{ab},
\end{equation}
where $H_{0} =\hbar \omega _{0}\sigma _{a}^{+}\sigma _{a}^{-}+\hbar \omega_{0}\sigma _{b}^{+}\sigma _{b}^{-}+\sum_{k}\hbar \omega _{k}b_{k}^{\dagger}b_{k}$ are the Hamiltonian of the quantum system, the ancillary system and the environment, $H_{\text{I}} =\sum_{k}\hbar (g_{k}\sigma _{a}^{+}b_{k}+g_{k}^{\ast }\sigma_{a}^{-}b_{k}^{\dagger })+\sum_{k}\hbar (g_{k}\sigma _{b}^{+}b_{k}+g_{k}^{\ast }\sigma_{b}^{-}b_{k}^{\dagger })$ are the interaction Hamiltonian between the system-ancilla and the environment, $H_{ab} =\hbar J(\sigma _{a}^{+}\sigma _{b}^{-}+\sigma _{a}^{-}\sigma_{b}^{+})$ is the interaction Hamiltonian between the quantum system and the ancillary system.

The initial state of the system and the ancilla is assumed in the state $|eg\rangle $, and the initial state of the common environment is in the vacuum states $|0\rangle $. The evolved state in time $t$ is
\begin{equation}
|\psi _{\text{c}}(t)\rangle =A_{\text{c}}(t)|eg\rangle |0\rangle +B_{\text{c}}(t)|ge\rangle|0\rangle +\sum_{k}C_{\text{c}}^{k}(t)|gg\rangle |1_{k}\rangle .\label{eq:psic}
\end{equation}

Inserting the state $|\psi _{c}(t)\rangle $ into the Schr\"{o}dinger equation, one can obtain the following differential equations
\begin{subequations}
\begin{align}
i\dot{A}_{\text{c}}(t)& =\omega _{0}A_{\text{c}}(t)+JB_{c}(t)+\sum_{k}g_{k}C_{\text{c}}^{k}(t), \\
i\dot{B}_{\text{c}}(t)& =\omega _{0}B_{\text{c}}(t)+JA_{c}(t)\sum_{k}g_{k}C_{\text{c}}^{k}(t), \\
i\dot{C}_{\text{c}}^{k}(t)& =\omega _{k}C_{\text{c}}^{k}(t)+g_{k}^{\ast}A_{\text{c}}(t)+g_{k}^{\ast }B_{\text{c}}(t).
\end{align}
\end{subequations}
Using the Laplace transformation and its inverse, the coefficient $A_{\text{c}}(t)$ in Eq. (\ref{eq:psic}) can be solved analytically
\begin{equation}
A_{\text{c}}(t)=a_{\text{c}}(t)e^{-i\omega _{0}t},
\end{equation}
where $a_{\text{c}}(t)=\frac{1}{2}(e^{iJt}+e^{-\frac{1}{2}(\lambda +iJ)t}h_{\text{c}})$ with the coefficients $h_{\text{c}}=\cos [\frac{1}{2}d_{\text{c}}t]+\frac{\lambda -iJ}{d_{\text{c}}}\sinh [\frac{1}{2}d_{\text{c}}t]$ and $d_{\text{c}}=\sqrt{-J^{2}-2iJ\lambda +\lambda(\lambda-4\gamma _{0})}$.

\begin{figure}[t]
\centering
\includegraphics[width=0.9\columnwidth]{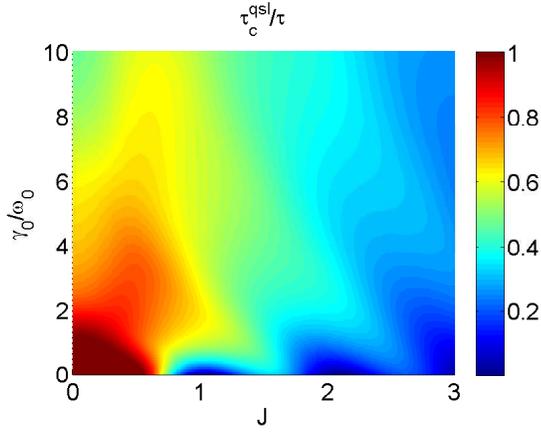}\newline
\caption{The contour of ratio $\tau_{\text{c}}^{\text{qsl}}/\tau$ as the function of environment coupling parameter $\gamma_0$ and the hopping interaction parameter $J$. The coupled system-ancilla interact with common environment, the spectral width parameter is chosen as $\lambda=2$ (in units of $\omega_0$) and the actual driving time is $\tau=3$.}\label{fig2}
\end{figure}

The ML bound of quantum speed limit time for the system of interest based on the operator norm is given as
\begin{align}
\tau_{\text{c}}^{\text{qsl}}&=\frac{\sin^2[\mathcal{L}(\rho_0,\rho_{\tau})]}{\Lambda_{\tau}^{\text{op}}}  \notag \\
&=\frac{1-\vert A_{\text{c}}(\tau)\vert^2}{\frac{1}{\tau}\int_0^{\tau}2\vert\text{Re}[\dot{A}_{\text{c}}(t)A_{\text{c}}^*(t)]\vert dt}.  \label{eq:cqsl}
\end{align}
The ratio between the quantum speed limit time and the actual driven time $\tau_{\text{c}}^{\text{qsl}}/\tau$ as the function of environment coupling strength $\gamma_0$ and hopping interaction $J$ is given in Fig. \ref{fig2}, the spectral width parameter is $\lambda =2$ (in units of $\omega _{0}$), and the actual driven time is chosen as $\tau =3$, which the parameters are chosen as the same with Fig.1.

The quantum speed limit time can be  significantly decreased along with the increasing of the environment coupling  parameters $\gamma_0$ and hopping interaction $J$, and demonstrates stronger periodic oscillation phenomenon. The similar property can also be found in the dynamics of the quantum correlation \cite{Wu18aop} based on the skew information \cite{Girolami13,Wu14}. For the dynamics of the system, one should notice that it can not be reduced to the damped Jaynes-Cummings model when the hopping interaction $J$ is zero. Because of the common interaction mechanism between the system-ancilla and environment, the excited state of the system can be decayed to the ground state and emit a photon into the environment, while this photon may be simulated to the ancilla. It is maybe the reason why the quantum speed limit time (\ref{eq:cqsl}) of the system is decreased markedly compared with the independent environment case (\ref{eq:2qsl}).

\section{Only the system of interest interacting with environment}

In this section, we will consider what will happen to the quantum speed limit time when only the system of interest interacts with the environment. The Hamiltonian of the system-ancilla and environment is
\begin{equation}
H_1=H_{a}+H_{b}+H_{ab},
\end{equation}
where $H_{a} =\hbar \omega _{0}\sigma _{a}^{+}\sigma _{a}^{-}+\sum_{k}\hbar \omega_{k}a_{k}^{\dagger }a_{k}+\sum_{k}\hbar (g_{k}^{\ast }\sigma _{a}^{+}a_{k}+\text{h.c})$ are the Hamiltonian of the quantum system interaction with the environment, $H_{b} =\hbar \omega _{0}\sigma _{b}^{+}\sigma _{b}^{-}$ is the Hamiltonian of the ancillary system, and $H_{ab} =\hbar J(\sigma _{a}^{+}\sigma _{b}^{-}+\sigma _{a}^{-}\sigma_{b}^{+})$ is the interaction Hamiltonian between the quantum system and the ancillary system.

The initial state is at $|eg\rangle |0\rangle _{a}$, i.e., the system is at the excited state $|e\rangle $ and its surrounding environment is at the vacuum state $\vert0\rangle$, and the ancilla is at the ground state $|g\rangle$. The evaluated state in time $t$ is
\begin{align}
|\psi _{1}(t)\rangle =& A_{1}(t)|eg\rangle |0\rangle _{a}+B_{1}(t)|ge\rangle|0\rangle _{a}  \notag \\
& +\sum_{k}C_{1}^{k}(t)|gg\rangle\vert1_{k}\rangle _{a}.  \label{eq:1state}
\end{align}

Utilizing the similar derivative processes in Eqs. (\ref{eq:2rho}-\ref{eq:2qsl}), the ML bound of quantum speed limit time based on the operator norm can be arrived as follows
\begin{align}
\tau _{1}^{\text{qsl}}& =\frac{\sin ^{2}[\mathcal{L}(\rho _{0},\rho _{\tau})]}{\Lambda _{\tau }^{\text{op}}}  \notag \\
& =\frac{1-|A_{1}(\tau )|^{2}}{\frac{1}{\tau }\int_{0}^{\tau }|\text{Re}[\dot{A}_{1}(t)A_{1}^{\ast }(t)]|dt}.  \label{eq:1qsl}
\end{align}

Substituting the state $|\psi _{1}(t)\rangle $ into the Schr\"{o}dinger equation, one can obtain the differential equations as
\begin{subequations}
\begin{align}
i\dot{A}_{1}(t)& =\omega _{0}A_{1}(t)+JB_{1}(t)+\sum_{k}g_{k}C_{1}^{k}(t), \\
i\dot{B}_{1}(t)& =\omega _{0}B_{1}(t)+JA_{1}(t), \\
i\dot{C}_{1}^{k}(t)& =\omega _{k}C_{1}^{k}(t)+g_{k}^{\ast }A_{1}(t).
\end{align}
\end{subequations}
Using the Laplace transformation and its inverses, the coefficient $A_{1}(t)$ in quantum speed limit time (\ref{eq:1qsl}) is
\begin{equation}
A_{1}(t)=a_{1}(t)e^{-i\omega _{0}t},
\end{equation}
where
\begin{equation}
a_{1}(t)=\sum_{i\neq j\neq k}\frac{e^{u_{i}t}u_{i}(\lambda +u_{i})}{(u_{i}-u_{j})(u_{i}-u_{k})}\label{a1}
\end{equation}
with $u_{i}$  the solutions of the cubic equation $2s^{3}+2\lambda s^{2}+(2J^{2}+\gamma _{0}\lambda)s+2J^{2}\lambda =0$.

\begin{figure}[t]
\centering
\includegraphics[width=0.9\columnwidth]{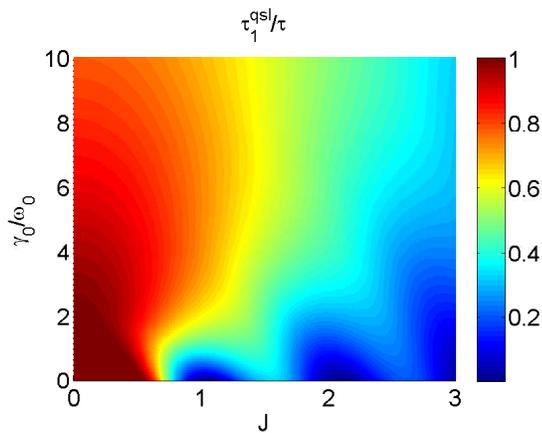}\newline
\caption{The ratio $\tau_1^{\text{qsl}}/\tau$ as the function of environment coupling parameter $\gamma_0$ and the system-ancilla hoppling strength $J$. Only the system of interest interacts with environment, and the spectral width parameter is chosen as $\lambda=2 $ (in units of $\omega_0$), and the actual driven time is $\tau=3$.}\label{fig3}
\end{figure}

The ratio between the quantum speed limit time and the actual drive time $\tau _{1}^{\text{qsl}}/\tau $ as the function of environment coupling strength parameter $\gamma _{0}$ and the hopping interaction coefficient $J$ is given in Fig. \ref{fig3}. The other parameters are chosen as the same with that in Figs. \ref{fig1} and \ref{fig2}, i.e., the spectral width parameter is chosen as $\lambda =2$ (in units of $\omega_{0}$), and the actual drive time is $\tau =3$. The quantum speed limit time in Eq. (\ref{eq:1qsl}) shows the similar properties to QSLT in Eq. (\ref{eq:2qsl}), however, the quantum speed limit time in Eq. (\ref{eq:2qsl}) is tighter than the one in Eq. (\ref{eq:1qsl}). Compared with the condition that coupled system-ancilla interact with independent environments, Fig. \ref{fig3} indicates that the evolution of the system can be accelerated faster with the assistance of the ancillary system when only the system of interest interacts with environment. If the hopping interaction $J$ between the system and ancilla strength is 0, the coefficient $a_{1}(t)$ in Eq. (\ref{a1}) can be simplified as $a_{1}(t)=e^{-\frac{\lambda t}{2}}(\cosh [\frac{\text{d}t}{2}]+\frac{\lambda }{\text{d}}\sinh [\frac{\text{d}t}{2}])$ with $\text{d}=\sqrt{\lambda (\lambda -2\gamma _{0})}$, which is the solution of the damped Jaynes-Cummings model.

\section{Conclusion and discussion}

When the quantum system interaction with the ancillary system through the hopping interaction in different non-Markovian environments, the behaviour of the quantum speed limit time for the quantum system exhibit similar feature, however, the degree that the evolution can be sped up is different  due to the interaction structure with environment.  In generally, for the case that the coupled system-ancilla interacting with common environment, the quantum speed limit time is the shortest one. When the initial state of the system-ancilla is $\vert eg\rangle$, in the evolution of the quantum information processing, there are two ways that the excited state of the system can be decay into the ground state, one is the interaction between the system and the common environment, and another is the hopping interaction between the system and the ancilla. Due to the interaction between the ancilla and the common environment, the radiated photon can also  be exchanged between the system and the ancilla. While, for the case that the coupled system-ancilla interacting with independent environment, the quantum speed limit time is the longest one. Different with the common environment, the radiated photon can not be exchanged between the system and ancilla through the independent environment. For the case  that only the system of interest interacting with environment, one can give a similar analysis.

In summery, the effects of ancillary system on the quantum speed limit in different non-Markovian environments were investigated. The quantum system and the ancillary system are coupled through the hopping interaction $\hbar J(\sigma _{a}^{+}\sigma _{b}^{-}+\sigma _{a}^{-}\sigma_{b}^{+})$, and we studied the case that the coupled two-level system-ancilla interact with independent environments, common environment, and only the quantum system of interest interacts with environment, respectively. In all of those models, the quantum speed limit time becomes shorter along with the enhancement of coupling interaction parameter$\gamma_0$ between the system-ancilla and environments, and the QSLT shows a periodic oscillation phenomenon along with the hopping interaction parameter$J$ increasing. Our results indicate that the hopping interaction between the quantum system and the ancillary system and the interaction mechanism between the system-ancilla and environment determine the degree of the quantum system evolution can be accelerated.

\section*{Acknowledgments}

 This work was supported by the Scientific and Technological Innovation Programs of Higher Education Institutions in Shanxi under Grant No. 2019L0527. Yu was supported by the National Natural Science Foundation of China under Grant No. 11775040.


\begin{thebibliography}{99}
\bibitem{Mandelstam45} Mandelstam, L., Tamm, I.: The uncertainty relation between energy and time in nonrelativistic quantum mechanics.  J. Phys. (USSR) \textbf{9}, 249 (1945)

\bibitem{Margolus98} Margolus, N., Levitin, L. B.: The maximum speed of dynamical evolution. Phys. D \textbf{120}, 188 (1998)

\bibitem{Levitin09} Levitin, L. B., Toffoli, T.: Fundamental llimit on the rate of quantum dynamics: the unified bound is tight. Phys. Rev. Lett. \textbf{103}, 160502 (2009)
\bibitem{Fleming73} Fleming, G. N.: A unitarity bound on the evolution of nonstationary states. Nuovo Cimento. \textbf{16}, 232 (1973)
\bibitem{Bhattacharyya83} Bhattacharyya, K.: Quantum decay and the Mandelstam-Tamm time-energy inequality. J. Phys. A \textbf{16}, 2993 (1983)
\bibitem{Anandan90} Anandan, J., Aharonov, Y.: Geometry of quantum evolution. Phys. Rev. Lett. \textbf{65}, 1697 (1990)
\bibitem{Pati91} Pati, A. K.: Relation between phases and distance in quantum evolution. Phys. Lett. A \textbf{159}, 105 (1991)
\bibitem{Vaidman92} Vaidman, L.: Minimum time for the evolution to an orthogonal quantum state. Am. J. Phys. \textbf{60}, 182 (1992)
\bibitem{Brody03} Brody, D. C.: Elementary derivation for passage times. J.Phys. A: Math. Gen. \textbf{36}, 5587 (2003)
\bibitem{Jones10} Jones, P. J., Kok, P.:  Geometric derivation of the quantum speed limit. Phys. Rev. A \textbf{82}, 022107 (2010)
\bibitem{Campaioli18} Campaioli, F., Pollock, F. A., Binder, F. C., Modi, K.: Tightening quantum speed limits for almost all states. Phys. Rev. Lett. \textbf{120}, 060409 (2018)

\bibitem{Caneva09} Caneva, T., Murphy, M., Calarco, T., Fazio, R., Montangero, S., Giovannetti, V., Santoro, G. E.: Optimal control at the quantum speed limit. Phys. Rev. Lett. \textbf{103}, 240501 (2009)
\bibitem{Bekenstein81} Bekenstein, J. D.: Energy cost of information transfer. Phys. Rev. Lett. \textbf{46}, 623 (1981)

\bibitem{Breuer16} Breuer, H. P., Laine, E. M., Piilo, J., Vacchini, B.: Colloquium: non-Markovian dynamics in open quantum systems. Rev. Mod. Phys. \textbf{88}, 021002 (2016)

\bibitem{Taddei13} Taddei, M. M., Escher, B. M., Davidovich, L., de Matos Filho, R. L.: Quantum speed limit for physical processes. Phys. Rev. Lett. \textbf{110}, 050402 (2013)
\bibitem{Campo13} del Campo, A., Egusquiza, I. L., Plenio, M. B., Huelga, S. F.: Quantum speed limit in open system dynamics.  Phys. Rev. Lett. \textbf{110}, 050403 (2013)
\bibitem{Deffner13} Deffner, S., Lutz, E.: Quantum speed limit for non-Markovian dynamics. Phys. Rev. Lett. \textbf{111}, 010402 (2013)

\bibitem{Frey16} Frey, M. R.: Quantum speed limits-primer, perspectives, and potential future directions. Quantum Inf. Process. \textbf{15}, 3919 (2016)
\bibitem{Deffner17jpa} Deffner, S., Campbell, S.: Quantum speed limits: from Heisenbergs uncertainty principle to optimal quantum control. J. Phys. A: Math. Theor. \textbf{50}, 453001 (2017)


\bibitem{Xu14} Xu, Z. Y., Luo, S., Yang, W. L., Liu, C., Zhu, S.: Quantum speedup in a memory environment. Phys. Rev. A \textbf{89}, 012307 (2014)
\bibitem{Zhang14} Zhang, Y. J., Han, W., Xia, Y. J., Cao, J. P., Fan, H.: Speedup of quantum evolution of multiqubit entanglement states.  Sci. Rep. \textbf{4}, 4890 (2014)
\bibitem{Wu15} Wu, S. X., Zhang, Y., Yu, C. S., Song, H. S.: The initial-state dependence of the quantum speed limit. J. Phys. A: Math. Theor. \textbf{48}, 045301 (2015)
\bibitem{Zhang15} Zhang, Y. J., Han, W., Xia, Y. J., Cao, J. P., Fan. H.: Classical-driving-assisted quantum speed-up. Phys. Rev. A \textbf{91}, 032112 (2015)
\bibitem{Sun15} Sun, Z., Liu, J., Ma, J., Wang, X.: Quantum speed limits in open systems: non-Markovian dynamics without rotating-wave approximation. Sci. Rep. \textbf{5}, 8444 (2015)
\bibitem{Liu16} Liu, H. B., Yang, W. L., An, J. H., Xu, Z. Y.: Mechanism for quantum speedup in open quantum systems. Phys. Rev. A \textbf{93}, 020105 (2016)
\bibitem{Song16} Song, Y. J., Kuang, L. M., Tan, Q. S.: Quantum speedup of uncoupled multiqubit open system via dynamical decoupling pulses. Quantum Inf. Process. \textbf{15}, 2325 (2016)
\bibitem{Ekt17} Ektesabi, A., Behzadi, N., Faizi, E.: Improved bound for quantum-speed-limit time in open quantum systems by introducing an alternative fdelity. Phys. Rev. A \textbf{95}, 022115 (2017)
 \bibitem{Wu18} Wu, S. X., Yu, C. S.: Quantum speed limit for a mixed initial state.  Phys. Rev. A \textbf{98}, 042132 (2018)
\bibitem{Awasthi20} Awasthi, N., Haseli, S., Johri, U. C., Salimi, S., Dolatkhah, H., Khorashad, A. S.: Quantum speed limit time for correlated quantum channel. Quantum Infor. Process. \textbf{19}, 10 (2020)
\bibitem{Wu20sr} Wu, S. X., Yu. C. S.: Quantum speed limit based on the bound of Bures angle. Sci. Rep. \textbf{10}, 5500 (2020)
\bibitem{Jing16} Jing, J., Wu, L. A., del Campo, A.: Fundamental speed limits to the generation of quantumness. Sci. Rep. \textbf{6}, 38149 (2016)

\bibitem{Pires16} Pires, D. P., Cianciaruso, M., C\'{e}leri, L. C., Adesso, G., Soares-Pinto, D. O.: Generalized geometric quantum speed limits. Phys. Rev. X \textbf{6}, 021031 (2016)
\bibitem{Mar16} Marvian, I., Spekkens, R. W., Zanardi, P.: Quantum speed limits, coherence, and asymmetry. Phys. Rev. A \textbf{93}, 052331 (2016)

\bibitem{Deffner19} Deffner, S.: Quantum speed limits and the maximal rate of information production. Phys. Rev. Res. \textbf{2}, 013161 (2019)
\bibitem{Campaioli20} Campaioli, F., Yu, C. S., Pollock, F. A., Modi, K.: Resource speed limits: Maximal rate of resource variation. arXiv: 2004.03078

 \bibitem{Campbell17} Campbell, S., Deffner, S.: Trade-off between speed and cost in shortcuts to adiabaticity. Phys. Rev. Lett. \textbf{118}, 100601 (2017)
\bibitem{Xu18} Xu, Z. Y., You, W. L., Dong, Y. L., Zhang, C., Yang, W. L.: Generalized speed and cost rate in transitionless quantum driving. Phys. Rev. A \text{97}, 032115 (2018)
\bibitem{Demirplak08} Demirplak, M., Rice, S. A.: On the consistency, extremal, and global properties of counterdiabatic fields. J. Chem. Phys. \textbf{129}, 154111 (2008)
\bibitem{Takahashi13} Takahashi, K.: How fast and robust is the quantum adiabatic passage? J. Phys. A: Math. Theor. \textbf{46}, 315304 (2013)
\bibitem{Campo14} del Campo, A., Goold, J., Paternostro, M.: More bang for your buck: super-adiabatic quantum engines. Sci. Rep. \textbf{4}, 6208 (2014)
\bibitem{Funo17} Funo, K., Zhang, J. N., Chatou, C., Kim, K., Ueda, M., del Campo, A.: Universal work fluctuations during shortcuts to adiabaticity by counterdiabatic driving. Phys. Rev. Lett. \textbf{118}, 100602 (2017)

\bibitem{Sun19} Sun, S., Zheng, Y.: Distinct bound of the quantum speed limit via the gauge invariant distance. Phys. Rev. Lett. \textbf{123}, 180403 (2019)

\bibitem{Fogarty20} Fogarty, T., Deffner, S., Busch, T., Campbell, S.: Orthogonality catastrophe as a consequence of the quantum speed limit. Phys. Rev. Lett. \textbf{124}, 110601 (2020)
\bibitem{Puebla20} Puebla, R., Deffner, S., Campbell, S.: Kibble-Zurek scaling in quantum speed limits for shortcuts to adiabaticity. Phys. Rev. Res. 2, 032020 (2020)

\bibitem{GP19} Garc\'{i}a-Pintos, L. P., del Campo, A.: Quantum speed limits under continuous quantum measurements. New J. Phys. \textbf{21}, 033012 (2019)
\bibitem{Nicholson20} Nicholson, S. B., Garc\'{i}a-Pintos, L.P., del Campo, A., Green, J. R.: Time-information uncertainty relations in thermodynamics. Nature Phys. \textbf{16}, 1211 (2020)

\bibitem{Shanahan18} Shanahan, B., Chenu, A., Margolus, N., del Campo, A.: Quantum speed limits across the quantum-to-classical transition. Phys.Rev. Lett. \text{120}, 070401 (2018)
\bibitem{Okuyama18} Okuyama, M., Ohzeki, M.: Quantum speed limit is not quantum. Phys. Rev. Lett. \textbf{120}, 070402 (2018)
\bibitem{Deffner17njp} Deffner, S.: Geometric quantum speed limits: A case for Wigner phase space. New J. Phys. \textbf{19}, 103018 (2017)
\bibitem{Shiraishi18} Shiraishi, N., Funo, K., Saito, K.: Speed limit for classical stochastic processes, Phys. Rev. Lett. \textbf{121}, 070601 (2018)
\bibitem{Wu20cpb} Wu, S. X.,  Yu, C. S.: Margolus-Levitin speed limit across quantum to classical regimes based on trace distance. Chin. Phys. B  \textbf{29}, 050302 (2020)
\bibitem{Hu20} Hu, X., Sun S., Zheng Y.: Quantum speed limit via the trajectory ensemble. Phys. Rev. A \textbf{101}, 042107 (2020)

\bibitem{Horn85} R. A. Horn, and C. R. Johnson, Matrix Analysis (Cambridge: Cambridge University Press) (1985)

\bibitem{Wu18aop} Wu, S. X., Zhang, Y., Yu, C. S.: Local quantum uncertainty guarantees the measurement precision for two coupled two-level systems in non-Markovian environment. Annals of Physics \textbf{390}, 71 (2018)
\bibitem{Girolami13} Girolami, D., Tufarelli, T., Adesso G. Characterizing nonclassicalcorrelations via local quantum uncertainty. Phys. Rev. Lett. \textbf{110}, 240402 (2013)
\bibitem{Wu14} Wu, S. X., Zhang, J., Yu., C. S., Song, H. S.: Uncertainty-induced quantum nonlocality. Phys. Lett. A \textbf{378}, 344 (2014)
\end{thebibliography}
\end{document}